\pdfoutput=1

\documentclass[acmsmall,nonacm]{acmart}

\makeatletter                   
\def\mdseries@tt{m}             
\makeatother                    

\usepackage{pervasives}
\usepackage{oxide}

\newcommand{\myparagraph}[1]{\paragraph{#1}}
\newcommand{\MIR}{\texttt{MIR}\xspace}
\newcommand{\lambdarust}{$\lambda_{\textrm{Rust}}$\xspace}
\renewcommand{\lang}{\texttt{Oxide}\xspace}
\renewcommand{\langl}[1]{\texttt{Oxide}$_{#1}$\xspace}

\newcommand{\mli}[1]{\mintinline{ocaml}{#1}}

\setcopyright{none}

\begin{document}

\title{Rust Distilled: An Expressive Tower of Languages}


\author{Aaron Weiss} \affiliation{\institution{Northeastern University}}
\affiliation{\institution{Inria Paris}}

\author{Daniel Patterson} \affiliation{\institution{Northeastern University}}

\author{Amal Ahmed} \affiliation{\institution{Northeastern University}}
\affiliation{\institution{Inria Paris}}

\begin{abstract}
  Rust represents a major advancement in production programming languages
  because of its success in bridging the gap between \emph{high-level}
  application programming and \emph{low-level} systems programming. At the heart
  of its design lies a novel approach to \emph{ownership} that remains highly
  programmable.

  In this talk, we will describe our ongoing work on designing a formal
  semantics for Rust that captures ownership and borrowing without the details
  of lifetime analysis. This semantics models a high-level understanding of
  ownership and as a result is close to source-level Rust (but with full type
  annotations) which differs from the recent RustBelt effort that essentially
  models \MIR, a CPS-style IR used in the Rust compiler. Further, while RustBelt
  aims to verify the safety of \rusti{unsafe} code in Rust's standard library,
  we model standard library APIs as primitives, which is sufficient to reason
  about their behavior. This yields a simpler model of Rust and its type system
  that we think researchers will find easier to use as a starting point for
  investigating Rust extensions. Unlike RustBelt, we aim to prove type soundness
  using \emph{progress and preservation} instead of a Kripke logical relation.
  Finally, our semantics is a family of languages of increasing \emph{expressive
    power}, where subsequent levels have features that are impossible to define
  in previous levels. Following Felleisen, expressive power is defined in terms
  of \emph{observational equivalence}. Separating the language into different
  levels of expressive power should provide a framework for future work on Rust
  verification and compiler optimization.
\end{abstract}

\maketitle

\section{Introduction}

Programming languages have long been divided between ``systems'' languages,
which enable low-level reasoning that has proven critical in writing systems
software, and ``high-level'' languages, which empower programmers with
high-level abstractions to write software more quickly and more safely. For many
language researchers then, a natural goal has been to try to enable both
low-level reasoning and high-level abstractions in one language. To date, the
Rust programming language has been the most successful endeavour toward such a
goal.

Nevertheless, Rust has also developed something of a reputation for its
complexity amongst programmers. It would seem almost every new Rust programmer
has their own tale of \emph{fighting the borrow checker} with its own mess of
unfamiliar type errors and associated stress. A natural question to wonder then
is if this reality is inevitable. We argue it is not! The challenge of learning
Rust is a familiar one---namely, learning new semantics is \emph{hard}. While
analogies by syntax make some aspects of Rust more comfortable to imperative
programmers, one cannot escape having to understand the novel semantics of
ownership in Rust, and for new programmers, it is tempting to get caught up in
the details of lifetime inference and analysis. While these details are
important for building an \emph{efficient} analysis, we feel they are
inappropriate for building a high-level mental model of the \emph{meaning} of
ownership, and hope that intuitions gleaned from our semantics can help. Of
course, we don't anticipate that beginner programmers will work through the
semantics directly. Instead, we believe that semanticists (in this case,
ourselves) have a secondary role as teachers---to distill semantic intuitions
into simple, clear explanations.

While there are some existing formalizations of Rust, we believe that none of
them are sufficient for our goals of (1)~understanding ownership as a seasoned
Rust programmer does and (2)~reasoning about how abstractions that rely on
\rusti{unsafe} code---such as those provided by the standard library---affect
the language's expressivity. The first major effort came in the form of
Patina~\cite{reed15:patina}, a formalization of an early version of Rust with
partial proofs of progress and preservation. More recently, the most well-known
and complete effort in formalizing Rust is RustBelt~\cite{jung18} whose
\lambdarust has already proven useful in verifying that major pieces of
\rusti{unsafe} code in the standard library do not violate Rust's safety
guarantees. Nevertheless, the low-level nature of \lambdarust as a language in
continuation-passing style makes it harder to use for source-level reasoning.
Also, RustBelt's goal of \emph{verifying} the \rusti{unsafe} code in Rust's
standard library means that \lambdarust has a much more complex type system and
lifetime logic than is necessary for \emph{understanding} ownership and
borrowing.

\section{Formalizing Rust}

In our talk, we will describe work in progress on developing \lang, a formal
semantics that aims to capture the essence of Rust with inspiration from linear
capabilities~\cite{ahmed06:linrgn} and fractional
permissions~\cite{boyland03:fractional}. To understand the core principles of
how we model ownership and borrowing in our semantics, it is helpful to look at
a simple example in Rust with its corresponding form in \lang. This example
declares a binding, and then \emph{immutably borrows} it.

\begin{center}
  \begin{minipage}{.18\textwidth}
\begin{minted}{rust}
let x = 5;
let y = &x;
\end{minted}
  \end{minipage}
\end{center}

In \lang, our code remains largely the same, but we make stack allocation
explicit via the \rusti{alloc} operator, and insert the usage of \rusti{drop}
that Rust would ordinarily infer. We also include annotations naming the
\emph{regions} that are being created (when we \rusti{alloc} or \rusti{borrow})
and destroyed (when we \rusti{drop}). To aid in comprehension, we also include
comments describing the state of an important static context as it changes
during type checking.

\begin{center}
  \begin{minipage}{.7\textwidth}
\begin{minted}{rust}
// P = {}
let imm x = alloc 'x 5;
// P = { 'x $\mapsto$ (u32, 1, {}) }
let imm y = borrow imm 'y x;
// P = { 'x $\mapsto$ (u32, $^1/_2$, {}), 'y $\mapsto$ (u32, $^1/_2$, { $\epsilon$ $\mapsto$ 'x }) }
drop 'y;
// P = { 'x $\mapsto$ (u32, 1, {}) }
drop 'x;
// P = {}
\end{minted}
  \end{minipage}
\end{center}

\noindent In particular, these comments describe the state of our region context
(denoted $\oxrenv$) after type checking each expression. This context contains a
mapping from region names \rusti{'r} to a triple of the region's type, its
fractional capability, and some additional metadata. We can see on line 3 that
when allocating a new region \rusti{'x} for a numeric constant, we associate it
with its type \rusti{u32} (an unsigned 32-bit integer), a whole capability
(denoted $\textbf{1}$), and no additional metadata. Then, when we borrow
immutably from \rusti{x} on line 4, we create a new region \rusti{'y} that takes
half of the capability and records that it is aliased from the region
\rusti{'x}. This metadata about aliasing is then used on line 6 to return the
half-capability to \rusti{'x} when we drop \rusti{'y}. This sort of automatic
management is a departure from typical presentations of linear
capabilities---where they are instead first-class values which are threaded
manually through the program---but more closely resembles the programming style
of Rust. Finally, note that dropping \rusti{'x} on line 8 corresponds to
different operational behavior than dropping \rusti{'y} on line 6. Since we have
a full capability for \rusti{'x} on line 8 and since there is no metadata
indicating that we must return the capability to some other region,
operationally this situation corresponds to freeing the data on the stack.

It is also important to note the departures from Rust in the wild. Specifically,
to have a capability guard the use of each value, it must be associated with a
region (since capabilities are always tied to regions). Thus, in \lang, all
values are used under references. One view of this model is that the mandatory
reference makes explicit the notion that the value is placed somewhere on the
stack. Further, this decision enables us to simplify our model by treating
\emph{moves} as \emph{mutable borrows} since both require full ownership
represented by a whole capability.

In the rest of this section, we discuss our plans for formalizing \lang as a
tower of languages and then give more detail about our current model.

\subsection{A Tower of Languages}

Though we introduced it as a single language, \lang is actually a \emph{family}
of languages that capture increasing levels of expressive
power~\cite{felleisen90}. The language we've already seen above represents
``safe Rust'' without any features from the standard library---we call this
\langl{0}. Subsequent language levels \langl{n+1} are achieved by extending each
language \langl{n} with abstractions (functionality) implemented using
\rusti{unsafe} code. We move up a language level, saying that \langl{n+1} is
more expressive than \langl{n}, when there exist observationally equivalent
programs in \langl{n}, that are \emph{not} observationally equivalent in
\langl{n+1}.\footnote{Observational equivalence for each level, \langl{n},
is defined as the standard notion of contextual equivalence for that language.}
We say \langl{n+1} is "more expressive" than \langl{n} since
\langl{n+1} has contexts with greater power that allows them to tell apart
programs that cannot be distinguished by contexts in \langl{n}.

This model of Rust as a family of languages at different levels of expressive
power gives us a way of precisely talking about what code refactoring, compiler
optimization, and program reasoning is justified given our codebase and
assumptions about the language level of code we link with. In particular, we can
say for real Rust code---which might contain some \rusti{unsafe} blocks---precisely what \rusti{unsafe} abstractions have been considered, giving us a way
to reason about observational equivalence of Rust programs.

\myparagraph{Allocation on the Heap} In \langl{1}, we extend \langl{0} with
\rusti{Vec<T>} which increases the expressivity of our language by giving us
access to the heap. Readers familiar with Rust might note that \rusti{Box<T>} is
typically thought of as the ``heap-allocated type'', but we chose \rusti{Vec<T>}
because it is more general (a \rusti{Box} is a \rusti{Vec} of length $1$).
Further, in principle, \rusti{Vec} alone is sufficient to write interfaces
observationally equivalent to data structures from \rusti{std::collections} like
\rusti{HashMap}, \rusti{BTreeMap}, and \rusti{BinaryHeap}---assuming, as is
typical, that performance is not included in our notion of observation.

\myparagraph{Shared Memory with Rc} For \langl{2}, we include \rusti{Rc<T>} which
provides reference-counted pointers. Like immutable references, these pointers
can be used to share memory between different parts of the program, but unlike
immutable references, the information is tracked \emph{dynamically}. This
enables programs to recover mutable references at runtime when they know that
there are no additional aliases. It is this ability to recover mutable
references that raises the language's expressive power.

\myparagraph{RefCells for Interior Mutability} In \langl{3}, we include
\rusti{RefCell<T>} which provides a way for shared data to be mutated. This
capability is known in the Rust community as \emph{interior mutability} because
it is often used to hide manipulations of internal state to ultimately present
an immutable interface. Like with \rusti{Rc}, \rusti{RefCell} works by deferring
the necessary safety checks around mutation to runtime. Though not restricted to
the heap, \rusti{RefCell} is analogous to \mli{ref} in the ML tradition.

\myparagraph{Growing Further} Our current goal is to formalize \langl{0} through
\langl{3}. In the future, we could extend our family of
languages further, adding the ability to spawn threads (\langl{4}), communicate
between them (\langl{5}), and so on. Like the early levels, these extensions add
further complexity and increase language expressivity.

\subsection{A Further Look at \langl{0}}

With a high-level understanding of \lang in place, we can now take a closer look
at the core language, \langl{0}. It includes
allocation on the stack (denoted $\oxalloc{\oxcrgn}{\oxexpr}$), copying
($\oxcopy{\oxcrgn}{\oxexpr}$) and borrowing mutably
($\oxborrow{\oxcrgn}{\oxmut}{\oxid}$) and immutably
($\oxborrow{\oxcrgn}{\oximm}{\oxid}$), all of which create a fresh region bound
to $\oxcrgn$.  Note that in our formal syntax, we write $\oxcrgn$ for \rusti{'r}
seen in the earlier example and use $\oxmuta$ to collectively refer to
mutability quantifiers $\oxmut$ and $\oximm$.  \langl{0} also features let
bindings, assignment, branching, and pattern matching.  However, pattern matching
is restricted to allow only simple patterns---i.e. those without nesting and
\rusti{ref} patterns. \langl{0} also includes structs, tuples, enumerations,
and fixed-sized arrays with borrowing inside each data structure
($\oxborrow{\oxcrgn}{\oxmuta}{\oxid.\oxpath}$). Here, $\oxpath$ denotes the path
through the data structure, e.g. borrowing the first field of a tuple $\oxid$
is written $\oxborrow{\oxcrgn}{\oxmuta}{\oxid.0}$. Finally, \langl{0} requires all
drops to be explicit (denoted $\oxdrop{\oxcrgn}$).  Hence, with different
strategies for placing these drop expressions when we compile---in essence,
\emph{elaborate}---from Rust to \lang, we can model Rust both with and without the
upcoming non-lexical lifetimes feature.

\myparagraph{Type system} We make use of a type-and-effect system where effects
keep track of changes to the region context (written $\oxrenv$) caused by the
given expression. This takes the shape of a judgment of the form
\oxformtypjudgeplain, read ``in the global context $\oxdenv$ (which contains
structure definitions), type variable context $\oxkenv$, region context
$\oxrenv$, and variable context $\oxtenv$, $\oxexpr$ has type $\oxtype$ with
effect $\oxeff$.'' Our available effects include creating new regions
($\oxnewrgn{\oxcrgn}{\oxtype}{\oxfrac}{\oxpathset}$ where $\oxtype$ is the type
of the value stored in the new region $\oxcrgn$, $\oxfrac$ denotes the
fractional capability for the region, and $\oxpathset$ denotes the metadata seen
in the example), borrowing from one region into a new region
($\oxbrwrgn{\oxmuta}{\oxcrgn_1}{\oxcrgn_2}$), deleting a region
($\oxdelrgn{\oxcrgn}$), and updating (sub)regions during assignment
($\oxupdatergn{\oxcrgn_1}{\oxid}{\oxcrgn_2}$).

In order to prove type soundness using progress and preservation, we rely on
the usual trick of using an \emph{instrumented semantics}.  In our model, the
instrumented semantics maintains information about regions in order to track
aliasing/borrowing relationships in memory during runtime
(instead of maintaining a more traditional flat memory structure that only maps
locations to values).
Intuitively, the way to think about this is that, upon allocation (or copy), we
create a new region that essentially corresponds to what would normally be a
physical location, but when we borrow, we create fresh regions akin to ghost
locations that exist for the purpose of keeping track of the aliasing that
results from borrowing.  Nonetheless, both the physical and the ghost locations
are modeled as regions \rusti{'r}.

Concretely, our operational semantics is a relation on machine configurations
$\oxdynst{\oxvst}{\oxrst}{\oxexpr}$ where $\oxvst$ is a store that maps
variables to regions, $\oxrst$ is a region store that maps regions to the values
stored there---similar to the static environment $\oxrenv$ discussed earlier but
storing values instead of types---and $\oxexpr$ is the expression being
evaluated. The instrumented semantics contains enough information to enable a
proof via progress and preservation (though the proof is still in progress).  In
the future, we will provide an erasure procedure that, in essence, removes
information about ghost locations from our machine configurations, and prove an
operational correspondence between the instrumented semantics and the
post-erasure semantics.

\myparagraph{Current Status} At the time of this writing, we have specified
\langl{0} and \langl{1} and are working on the progress and preservation proofs
for \langl{0}.  We have a prototype type checker implemented in Scala for
experimentation, and the beginnings of a Coq formalization.
We plan to expand the Scala prototype further to include a compiler that
elaborates Rust programs into Oxide, as well as an interpreter for Oxide. This
will allow us to test our semantics against Rust for
accuracy (along the lines of~\citet{guha10:essence-js}'s testing of their core
calculus for JavaScript).

\section{A Rusty Future}

With a precise framework for reasoning about source-level Rust programs, we hope
that we, as a community, can build great things around Rust!  We already have a
number of ideas ourselves, many of which we are only just beginning to explore.

\myparagraph{Language Extensions} With semantics in hand, the eager programming
language researcher can jump at the opportunity to build nice, well-behaved
extensions to Rust. This can be useful in trying to evolve the language through
its RFC process~\cite{rust-rfcs} where informal formalisms have already begun to
crop up~\cite{ticki17:pi}. Meanwhile, \lang can also form the basis of
domain-specific extensions. For example, we are designing extensions for secure
multiparty computation~\cite{yao86:garbled-circuits, doerner17:oram} in the
style of Obliv-C~\cite{zahur15:oblivc}. Further, extensions can be built with
the particular focus of enabling Rust programmers to write more reliable and
correct software. This can include anything from verification-oriented language
features as in Liquid Haskell~\cite{vazou14:liquid-haskell} to tools for
symbolic execution~\cite{king76:symex} and beyond.

\myparagraph{Safe Interoperability} The Rust community has already begun to
recognize the importance of building higher-level interfaces for
interoperability with other programming languages~\cite{rust-neon,
  rust-curryrs}. We hope to use \lang to expand what is possible for these
interoperability frameworks. In particular, we want to build on prior work on
multi-language compilers~\cite{perconti14:fca, ahmed15:snapl} and linking
types~\cite{patterson17:snapl} to support \emph{provably safe} interoperation
between languages.

\myparagraph{Unsafe Code Guidelines} Finally, a pressing issue in the Rust
community remains the open question of what \rusti{unsafe} code is safe to
write~\cite{actix-web-unsafe}. With \lang, we believe we are laying
a foundation for answering such a question~\cite{matsakis16:levels}. Going
forward, we hope to use the intuitions from our work to contribute to the effort
to develop unsafe code guidelines.

\begin{acks}
  We wish to thank Niko Matsakis for his invaluable feedback, discussions, and
  blogging. We would also like to thank Denis Merigoux for his feedback on a
  draft of this paper. This work was done at Inria Paris during Fall 2017 and
  Spring 2018 while Amal Ahmed and Aaron Weiss were visiting the Prosecco team.
  This material is based upon work supported in part by the National Science
  Foundation under grants CCF-1453796 and CCF-1618732, and an NSF Graduate
  Research Fellowship (GRFP) for Aaron Weiss. This work is also supported in
  part by the European Research Council under ERC Starting Grant SECOMP
  (715753).
\end{acks}

\bibliography{amal}

\end{document}